\documentclass[showpacs,twocolumn,preprintnumbers,amsmath,amssymb]{revtex4}

\usepackage{graphicx}
\usepackage{dcolumn}
\usepackage{bm}

\begin{document}
\title{Resonator channel drop filters in a plasmon-polaritons metal}
\author{Sanshui Xiao}
\author{Liu Liu}
\author{Min Qiu}
\email{min@imit.kth.se} \affiliation{Laboratory of Optics,
Photonics and Quantum Electronics, Department of Microelectronics
and Information Technology, Royal Institute of Technology (KTH),
Electrum 229, 16440 Kista, Sweden}

\begin{abstract}
Channel drop filters in a plasmon-polaritons metal are studied. It
shows that light can be efficiently dropped. Results obtained by
the FDTD method are consistent with those from coupled mode
theory. It also shows, without considering the loss of the metal,
that the quality factor for the channel drop system reaches
$4000$. The quality factor decreases significantly if we take into
account the loss, which also leads to a weak drop efficiency.
\end{abstract}

 \maketitle

\section{Introduction}
Channel drop filters (CDFs), which single out one channel from
wavelength division multiplexed signals, and pass through other
channels undisturbed, are useful and essential elements for
photonic integrated circuits and dense wavelength division
multiplexing optical communication systems
\cite{Haus1992P1,Manolatou1999P2,Oda1991P1,Little1997P1}. Various
CDFs exist, such as fiber Bragg gratings, Fabry-Perot filters, and
arrayed waveguide gratings. Resonant CDFs, which involve
waveguide/cavity interaction, are attractive candidates for this
purpose because they can potentially realize the narrowest
linewidth for a given device size. In particular, resonant CDFs
implemented in photonic crystals can be made ultra-compact and
highly wavelength selective
\cite{Fan1998P1,Fan1998P2,Qiu2003P2,Kim2004P1,Qiu2005P1}.

A surface plasmon (SP) is a collective oscillation of the
electrons at the interface between a metal and a dielectric. SPs
give rise to surface-plasmon-waves, which are propagating
electromagnetic waves bound at the metal-dielectric interface
\cite{Rotman1951P1,Hurd1954P1,Elliott1954P1,Pendry2004P1}. A usual
dielectric waveguide cannot restrict the spatial localization of
optical energy beyond the $\lambda_0/2n$ limit, where $\lambda_0$
is the free space photon wavelength and $n$ is the refractive
index of the waveguide. As opposed to dielectric waveguides,
plasmonic waveguides have shown the potential to guide
subwavelength optical modes, the so-called surface plasmon
polaritons (SPP), at metal-dielectric interfaces, such as metallic
nanowires \cite{Weeber1999P1,Dickson2000P1} and metallic
nanoparticle arrays \cite{Quinten1998P1,Maier2003P1}. In this
letter, we investigate disk/ring resonator channel drop filters
realized in a two-dimensional (2D) plasmon-polaritons metal using
a finite-difference time-domain (FDTD) method \cite{TafloveFDTD}
with perfect matched layer boundary conditions, together with the
coupled mode theory.  A combination of FDTD techniques and Pade
approximation with Baker's algorithm is used for the quality
factor of the system and unload resonant cavity
\cite{Guo2001P1,Qiu2005P2}. In our numerical calculations, the
ring/disk resonator is described by a spatial discretization grid
in FDTD method, which naturally introduces a surface roughness. It
has been shown that the surface roughness leads to back
reflections into the counter propagating mode and a splitting of
the resonant peak \cite{Little1997P2}. Here, we use a spatial grid
size of $2.5nm$ in FDTD algorithm which we found to be sufficient
for the convergence of numerical results.

\begin{figure}[htb]
\centering\includegraphics[width=8 cm]{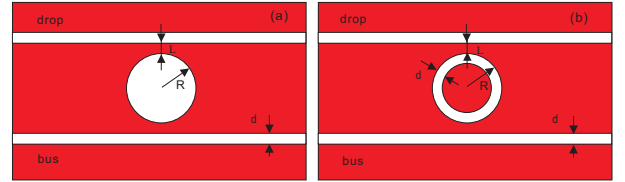}
\caption{Channel drop filter structure realized in a
plasmon-polaritons metal with (a) an air disk resonator; (b) a
ring resonator. }
\end{figure}
Consider firstly a resonant channel drop filter with an air
circular disk resonator in a plasmon-polaritons metal, where the
waveguide and the resonator are introduced by removing the metal
with a specific shape. The corresponding structure is shown in
Fig. 1 (a), where $d$, $L$ and $R$ is the waveguide width,
distance between the boundaries of the air disk and the waveguide,
and the radius of the air disk resonator, respectively. The radium
of the air disk is chosen as $1 \mu m$, and the dielectric
function of the metal (Silver) is described by the loss Drude
model:
\begin{eqnarray}
\varepsilon(\omega)=\varepsilon_\infty-\frac{(\varepsilon_0-\varepsilon_\infty)\omega_p^2}{\omega^2+2i\omega\nu_c},
\end{eqnarray}
where $\varepsilon_\infty$/$\varepsilon_0$ is the relative
permittivity at infinite/zero frequency, $\omega_p$ is the plasma
frequency, and $\nu_c$ is the collision frequency. We choose
$\varepsilon_\infty=4.017$, $\varepsilon_0=4.896$,
$\omega_p=1.419\times 10^{16}rad/s$ and $\nu_c=1.117\times
10^{14}rad/s$ for the Drude model, which fits the experimental
data quite well \cite{Palikbook}. The waveguide width, chosen as
$60 nm$, is much smaller than the wavelength so that only the
fundamental transverse magnetic ($TM$) waveguide mode is excited.
Our studies show, to realize a CDF system, that the $TM_0$ mode
waveguide  is preferred. The reason is that, compared with the
$TM_0$ mode waveguide, it is much difficult to couple the energy
from the waveguide to the resonator for the higher-order $TM$ mode
waveguide, which almost act as a perfect metallic waveguide.

Coupled mode or scattering theory are used to analyze
theoretically the interaction of a cavity resonator with a
waveguide system \cite{Manolatou1999P2,Xu2000P2}. But these works
consider waveguides with continuous translation symmetry and
ignore waveguide dispersion. Waks \emph{et. al.} derive an
analytical coupled mode theory including waveguide dispersion ,
especially for a photonic crystal waveguide \cite{Waks2005P1}.
However, for such a SPP waveguide, the conventional coupled mode
or scattering theory are also often suitable since the dispersion
for the SPP waveguide is quite weak in the frequency region of
interest, as well as for a optical fiber waveguide. According to
the coupled mode theory,  we can easily obtain the filter response
of the system \cite{Manolatou1999P2}
\begin{eqnarray}
&|R|^2&=\left|\frac{\sqrt{1/(\tau_{e1}\tau_{e2})}}{j(\omega-\omega_0)+1/\tau_o+2/\tau_e}\right
|^2,  \\
&|T|^2&=\left|1-\frac{1/\tau_{e1}}{j(\omega-\omega_0)+1/\tau_o+2/\tau_e}\right
|^2,  \\
&|D_L|^2&=\left|\frac{\sqrt{1/(\tau_{e1}\tau_{e4})}}{j(\omega-\omega_0)+1/\tau_o+2/\tau_e}\right
|^2,  \\
&|D_R|^2&=\left|\frac{\sqrt{1/(\tau_{e1}\tau_{e3})}}{j(\omega-\omega_0)+1/\tau_o+2/\tau_e}\right
|^2,
\end{eqnarray}
where $\omega_0$ is the resonant frequency, $1/\tau_o$ and
$1/\tau_e$ are the decay rate due to loss and the rate of decay
into the bus/drop waveguide, $R$ is the reflection from the input
port, $T$ is the transmission through the bus and $D_L$ and $D_R$
represent the transmission into the left and right ports of the
receiver. $1/\tau_{e1,3}$ and $1/\tau_{e2,4}$, defined as the
decay rates in the forward and backward direction, satisfy
\begin{eqnarray}
1/\tau_{e1}+1/\tau_{e2}=1/\tau_{e3}+1/\tau_{e4}=2/\tau_{e}.
\end{eqnarray}

In a travelling-wave mode, the power flows continuously in only
one direction in the resonator. It can be easily obtained that the
incident power in the bus in the forward direction is partially
transferred to the receiver in the backward direction, limited
only by the loss for the unloaded resonator, i.e., for
$1/\tau_0=0$ the channel with resonant frequency $\omega_0$, can
be transferred completely into the drop waveguide. For a pure
standing-wave mode, the resonant mode decays equally into the
forward and backward propagating waveguide mode. From Eqs.
(2)-(5), one knows that the maximum drop efficiency is
$|D_L|^2=|D_R|^2=0.25(1-\tau_e/\tau_o)$, i.e, half the input power
at resonant frequency can be dropped into the drop waveguide if
the loss for the unloaded resonator is ignored.

\begin{figure}[htb]
\centering\includegraphics[width=7 cm]{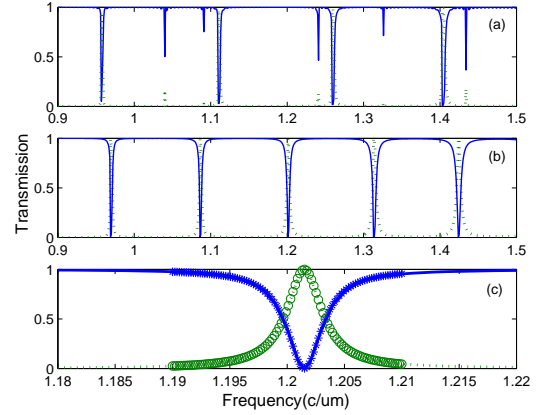}
\caption{Spectral response of the channel drop filter at the
forward bus (solid line) and backward drop waveguide (dotted line)
with (a) the air disk cavity; (b) the ring resonator cavity. The
circled/asterisk line in (c) represent the spectral response
obtained from the coupled mode theory and solid/dotted line
represent results from (b).}

\end{figure}

Look back to the channel drop filter system, as shown in Fig. 1
(a). Not only surface travelling resonant modes, but also standing
wave modes in the air disk resonator will be excited. Our
calculations show that the intrinsic quality factor ($Q$) for the
unload resonator is quite large and reaches $10^{10}$ if we ignore
the loss of the metal, and the quality factor ($Q_\|$) for the
filter system is about $600$ with the resonant frequency $f=1.261
(1/\mu m)$ (normalized to light velocity in vacuum $c$) when the
distance between the waveguide and resonator is $L=25nm$.
Basically, $Q_\|$ can increase if we enlarge $L$. Based on the
coupled mode analysis above, almost $100\%$ power can be dropped
by the surface-travelling mode, while the drop efficiency at one
output based on standing waves is quite small, at most $25\%$. The
spectral response after $500,000$ time steps, for $L=25nm$, are
shown in Fig. 2(a), where the solid and dotted line represent the
transmission at the forward bus and backward drop waveguide,
respectively. It can be seen from Fig. 2(a) that the light can be
efficiently dropped for some frequencies. For some resonant
frequencies, the drop efficiency is quite high, close to $100\%$,
due to the surface travelling mode. However, the drop efficiency,
governed by the standing wave, is below $25\%$, which is in
agreement with results from coupled mode theory. Figure 3(a) shows
the oscillation of the steady-state field distribution at a
resonant frequency with $f=1.404 (1/\mu m)$ , where almost all
energy is dropped by a surface travelling resonant wave.

\begin{figure}[htb]
\centering\includegraphics[width=7 cm]{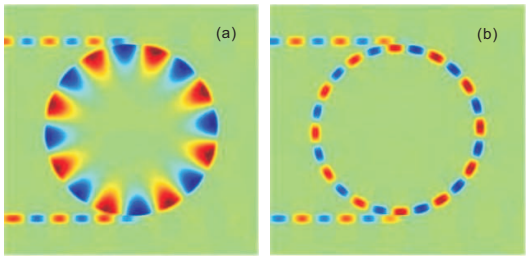}
\caption{Steady state $H_y$ field oscillation for a resonant
frequency of (a) $f=1.404 (1\mu m)$ ; (b) $f=1.426 (1\mu m)$.
Corresponding structures are shown in Fig. 1.}
\end{figure}

One important factor characterizing a channel drop filter is its
channel isolation ($CI$), defined as $CI=10log(P_1/P_2)$, where
$P_1$ and $P_2$ are the power of the selected channel transferred
to the drop waveguide and the power remaining in the bus
respectively. The channel isolation should be as large as possible
to avoid the cross talk. For the resonator with a surface
travelling mode mentioned above, $CI$ is quite large and
theoretically infinite if ignored the loss of the metal. However,
$CI$ for a single-mode standing wave resonator is quite small. It
is possible to get a high $CI$ for a single-mode travelling-wave
resonant filter using two standing-wave modes. In this letter, to
avoid exciting the standing-wave modes in the resonator, a channel
drop filter with a ring resonator is investigated, as shown in
Fig. 2(b). The width of the ring waveguide is same as the bus/drop
waveguide and the outer radius of the ring resonator is also
chosen as $1\mu m$. Mechanism in this system is quite similar to
the conventional ring resonator. Coupled mode theory tells us that
a single-mode travelling wave resonator side-coupled to the bus
and the receiver can fully transfer a channel at the resonant
frequency from the bus to the receiver, only limited by $Q_\|/Q$.

The spectral response at both the forward bus (solid line) and
backward drop waveguide (dotted line) are shown in Fig. 3(b) when
$L=25nm$. It can be seen from Fig. 3(b) that light can be almost
completely dropped for resonant frequencies. The results for the
resonant frequency $f=1.201 (1/\mu m)$, obtained by coupled mode
theory, are shown in Fig. 3(c) by circled and asterisk lines. It
can be seen from Fig. 3(c) that those obtained from coupled mode
theory are consistent with the results by FDTD method. Combined
with the resonant conditions ($N\times 2\pi=2\pi R n_{eff}k_0$),
the effective index ($n_{eff}$) for the ring waveguide with the
width of $60nm$, is about $1.339$ for the resonant frequency
$1.426 (1/\mu m)$ from the FDTD results. However, the effective
index for the straight waveguide with the same width is $1.501$
for the same resonant frequency. There is still no any good
analytical approximate method to calculate the effective index of
the ring waveguide when the working wavelength is close to the
radius of the ring. Figure 4(b) shows the oscillation of the
steady-state field distribution at a resonant frequency with
$f=1.426 (1/\mu m)$, where almost $100\%$ of the field energy is
transferred along the backward drop waveguide. However, the
quality factor for the filter is quite slow, only about $250$ for
the resonant frequency $f=1.426 (1/\mu m)$. Since the intrinsic
quality factor for the unload resonator is much high, the quality
factor of the channel drop filter can be increased if we enlarge
the distance between waveguide and ring resonator, which also
keeps a high drop efficiency. $Q_\|$ reaches $4\times 10^3$ when
$L=50 nm$.

A metal ($\varepsilon(\omega)=\varepsilon_{r}+i\varepsilon_{i}$)
is always a loss material especially in the visible and infrared
frequency region, which satisfies $\varepsilon_r<0$ and
$\mid\varepsilon_r\mid\gg\varepsilon_i$. The loss problem for a
straight SPP waveguide can be dealt with the first-order
perturbation theory, which considers the imaginary part of
permittivity as a perturbation for the real part of permittivity.
Consider a channel drop filter with a ring resonator in a lossy
plasmon-polaritons metal. If we take into account the loss of the
metal, the quality factor for the unloaded resonant cavity
decreases significantly. $Q$ is only about $60$ for the cavity
with a radius of $1\mu m$.  To increase the drop efficiency, we
shorten the distance between the waveguide and the cavity in order
to enlarge the coupling. We set the source and two detectors with
a distance of $0.5 \mu m$ with the center of the ring resonator in
the propagation direction. For the channel drop filter system with
$L=10nm$, the quality factor of the channel drop filter is only
$30$, which leads to weak drop efficiency. The spectral response
at both the forward bus (solid line) and backward drop waveguide
(dotted line) are shown in Fig. 4. It can be seen from Fig. 4 that
the drop efficiency is quite slow, as well as the quality factor.
The drop efficiency becomes larger with increasing the resonant
frequency. Since the quality factor for the filter becomes smaller
with increasing the resonant frequency, based on the coupled mode
theory, the drop efficiency will naturally increase.

\begin{figure}[htb]
\centering\includegraphics[width=6 cm]{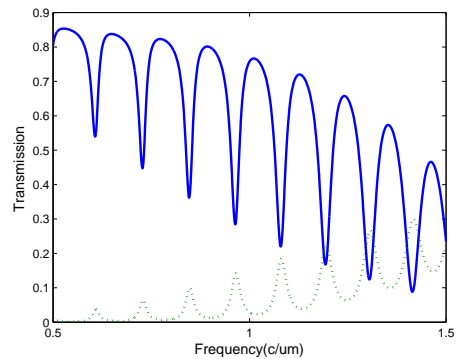}
\caption{Spectral response of the channel drop filter at the
forward bus (solid line) and backward drop waveguide (dotted line)
in a loss metal.}
\end{figure}

In Conclusion, we have investigated channel drop filters in a
plasmon-polaritons metal. Our results show that light can be
efficiently dropped in these channel drop systems. The results
obtained by FDTD method are consistent with the results from
coupled mode theory. Without considering the loss of the metal,
the quality factor of the channel drop systems reaches $4000$.
However, the quality factor decreases significantly if we take
into account  the loss of the metal. For the channel drop system
in a loss metal with $L=10 nm$, the maximum drop efficiency is
about $30\%$ and the quality factor is only $30$. Recently, many
theoretical and experimental results show that using surface
plasmon polaritons the scale of optoelectronic devices can be
shrunk by at least an order of magnitude \cite{Smolyaninov2005P1}.
People used dielectrics with gains to compensate the loss of the
metal \cite{Nezhad2004P1}. However, the loss of the
plasmon-polaritons metal is still a problem.

This work was supported by the Swedish Foundation for Strategic
Research (SSF) on INGVAR program, the SSF Strategic Research
Center in Photonics, and the Swedish Research Council (VR) under
Project No. 2003-5501.

\end{document}